\documentclass[journal=jctcce,manuscript=article]{achemso}

\usepackage{amsmath}
\usepackage{amssymb}

\newcommand{\kbt}{\mathrm{k_b}T}

\newcommand{\pos}{x}
\newcommand{\ppos}{(\pos)}

\newcommand{\pmf}{U_\mathrm{PMF}}
\newcommand{\pmfdensity}{p_\mathrm{PMF}}
\newcommand{\pmfbf}{\mathrm{e}^{- \beta U_\mathrm{PMF}}}
\newcommand{\pmfbfpp}{\mathrm{e}^{- \beta U_\mathrm{PMF} \ppos}}
\newcommand{\pmfpf}{Z_{\mathrm{PMF}}}

\newcommand{\ff}{U_\mathrm{FF}}
\newcommand{\ffdensity}{p_\mathrm{FF}}
\newcommand{\ffbf}{\mathrm{e}^{- \beta U_\mathrm{FF}}}
\newcommand{\ffbfpp}{\mathrm{e}^{- \beta U_\mathrm{FF} \ppos}}
\newcommand{\ffpf}{Z_{\mathrm{FF}}}

\newcommand{\deltau}{\Delta U}

\newcommand{\marg}{M}
\newcommand{\cond}{\eta}

\newcommand{\cost}{\mathcal{C}}

\author{Aleksander E. P. Durumeric}
\author{Gregory. A. Voth}
\email{gavoth@uchicago.edu}
\affiliation[University]
{Department of Chemistry, Chicago Center for Theoretical
Chemistry, James Franck Institute, and Institute for
Biophysical Dynamics, The University of Chicago, 5735 S.
Ellis Ave., Chicago, Illinois 60637, USA}

\title{Explaining classifiers to understand
coarse-grained models}

\begin{document}
\begin{tocentry}

Some journals require a graphical entry for the Table of Contents.
This should be laid out ``print ready'' so that the sizing of the
text is correct.

Inside the \texttt{tocentry} environment, the font used is Helvetica
8\,pt, as required by \emph{Journal of the American Chemical
Society}.

The surrounding frame is 9\,cm by 3.5\,cm, which is the maximum
permitted for  \emph{Journal of the American Chemical Society}
graphical table of content entries. The box will not resize if the
content is too big: instead it will overflow the edge of the box.

This box and the associated title will always be printed on a
separate page at the end of the document.

\end{tocentry}

\begin{abstract}
    Bottom-up coarse-grained molecular dynamics models are
    parameterized using complex effective
    Hamiltonians. These models are typically optimized to
    approximate high dimensional data from atomistic
    simulations. In contrast, human validation of these
    models is often limited to low dimensional statistics
    that do not necessarily differentiate between the CG
    model and said atomistic simulations.
    We propose that explainable
    machine learning can directly convey 
    high-dimensional error to scientists and use Shapley additive
    explanations  do so in two coarse-grained protein models.
\end{abstract}

\section{Introduction}

Atomistic molecular dynamics (MD) has provided scientific
insight into many
problems\cite{karplus2002molecular,shaw2010atomic,
buch2010high,lindorff2011fast,karplus2014significance}. 
Despite computational gains,
however, atomistic MD is still limited in terms of the time- and
space- scales it can access.
These limitations have
motivated the development of coarse-grained (CG) models that
simulate the original system at a minimal resolution, 
aiming to reduce computational cost while maintaining
quantitative
accuracy\cite{voth2008coarse,brini2013systematic,saunders2013coarse,noid2013perspective,
noid2013systematic,marrink2013perspective,potestio2014computer,pak2018advances,gkeka2020machine,
noe2020machine}.
A common class of CG models describe
molecular behavior using
MD at a coarser resolution than their atomistic counterparts; the behavior of these
simulations is then primarily controlled via an effective Hamiltonian.
For example, a
CG model may simulate a solvated protein by propagating
only the center of mass of each amino acid; the equilibrium distribution
could then be 
controlled by a Hamiltonian defined at the resolution of these
centers of mass.
 The behavior of
these models can be divided into their dynamic
and thermodynamic\footnote{Thermodynamic
here refers to long-time behavior related to the 
equilibrium distribution of the model. It includes both
issues related to estimating thermodynamic quantities, such
as pressure, and averages of functions of microstates.} properties.
While accurately reproducing the dynamics of a reference
atomistic system is an area of current
interest\cite{rudzinski2019recent}, the remainder of
this article focuses on thermodynamic issues, and
more specifically, the configurational distribution produced
by a CG model. We here limit our discussion to systems in the
canonical ensemble, and solely focus on the
configurational portion of this effective Hamiltonian, which we
refer to as the CG force-field.

There are many ways to create the effective Hamiltonian
characterizing a CG model\cite{voth2008coarse,brini2013systematic,
saunders2013coarse,noid2013perspective,
noid2013systematic,marrink2013perspective,potestio2014computer,
pak2018advances,gkeka2020machine,
noe2020machine}; these various approaches often
result in different effective force-fields.
Top-down methods aim to reproduce
observables that are coarser than the effective
Hamiltonian, such as partition
coefficients or interfacial tension. The coarse resolution of these
observables makes it possible to obtain reference data from either
experiment or simulation. In contrast, bottom-up
methods tend to require samples from a reference atomistic
simulation, which are mapped to the resolution of the CG
Hamiltonian and used as a target for
parameterization.\footnote{This distinction is complex; it
has been shown that behavior of various bottom-up techniques
can be implemented using low dimensional correlations, see
refs \cite{noid2007multiscale,shell2008relative}.} The variety 
of possible parameterization
techniques makes it valuable to understand how a proposed
force-field approximates a reference atomistic simulation at
the resolution of the CG model (i.e., that of the effective
Hamiltonian),
whether it be a reference simulation used to parameterize the model or one
created solely for external validation. 

However,
while a CG model is intrinsically coarser than the
atomistic model it represents, it is still high dimensional. For example,
the CG molecules in this paper are relatively
small but easily reach 36 dimensions, which is well beyond
what generic data visualizations (e.g., scatter plots or
histograms) can communicate to humans. These models are often
still amenable to visualization as groups of particles in 3D
space\cite{vmd} as the systems preferentially occupy a small portion
of the possible phase space (for example, a protein does not
dissociate into its constituent atoms--- its behavior is
strongly constrained by its primary and higher order
structure). However, while visual inspection can detect some
errors, it can be difficult to translate into
quantitative evaluations of the model. Established dimensional reduction
techniques\cite{ceriotti2019unsupervised,gkeka2020machine,noe2020machine} 
can be used to 
summarize the behavior present in the model and reference
data, but are not typically designed to find differences
between them.  In contrast, the original task of designing 
said force-fields does  
involve quantitative comparison of candidate
force-fields to reference data. In the case of 
top-down force-fields, optimization involves statistics that
are readily intepretable by computational scientists;
however, these  observables are
coarser than the resolution of the effective Hamiltonian. In
the case of bottom-up force-fields, while the considered
resolution is ideal, said optimization often
utilizes specific (and often opaque) computational
algorithms to optimize the Hamiltonian such that its high
dimensional configurational statistics approximate those of
a reference model. As a result, it can be difficult to 
understand how the behavior produced by a CG force-field
deviates from the ideal high dimensional behavior of the atomistic system,
even when such reference data is available.

This article will provide a way to compare two samples from
differing high dimensional free energy surfaces. The
analysis presented here does not specify a particular
process be used to generate these samples.  However, the
examples we will study have been created using bottom-up
coarse-graining techniques and we will borrow terminology
from the bottom-up coarse-graining literature to express our
ideas: for example, we will refer to the function which maps
each atomistic configuration to its CG
counterpart as the CG map, and we will refer to
the ideal effective force-field perfectly reproducing the
mapped atomistic statistics as the manybody potential of
mean force (manybody-PMF).  Furthermore, as the techniques we
describe provide a way to compare two samples from differing
high dimensional free energy surfaces, these approaches are
especially
pertinent to bottom-up parameterization strategies as a
mapped atomistic reference sample is readily available. 

As noted previously, the high dimensional nature of the data
produced by CG models makes it difficult to directly
visualize their full behavior.
When parameterizing atomistic models
using quantum mechanical data, computational scientists can
often additionally compare the energies or forces produced
by the reference method to those produced by the proposed
atomistic force-field, and use this configuration-wise error (or
atom-wise decompositions of this error) 
to study problematic areas of phase space (e.g. 
\citet{bartok2018machine}). This is
difficult to do with CG models as the point-wise evaluation
of the manybody-PMF is often not available and only a noisy
estimate may be available for the forces of the
manybody-PMF\footnote{It is possible to use constrained
fine-grained simulation to estimate the derivative of the
manybody-PMF\cite{stoltz2010free}.  The feasibility
of this approach depends on the system at hand and the level
of coarse-graining, but the results can 
analyzed using methods analogous to those described in this
article.}. An
attractive approach is to train multiple CG force-fields,
each with increasingly complex manybody interactions:
treating a higher order potential as if it were the true
manybody-PMF allows one to estimate the conditional free energy and
force differences at each configuration.  In this article we
show that a classifier can be used in lieu of training a
more complex CG force-field to obtain the same estimate of
force-field quality. While this alone allows the
computational scientist to isolate problematic
configurations based on force-field error, the larger
contribution of this work is the realization that the
classifier based approach, when combined with methods from
explainable ML/AI (collectively referred to as XAI in this
article), allows one to analyze the
force-field errors in a novel way.

XAI is a subfield of AI under active
development (for an overview of the corresponding
definitions, see refs
\cite{molnar2020book,arrieta2020explainable,
arya2019one,murdoch2019definitions,molnar2020interpretable,
makeproceedings}).
This issue of algorithmic transparency
is not new (see, e.g., refs
\cite{kodratoff1994,ruping2006learning}) however, as automated
decisions become more common in everyday life, an increasing
amount of scrutiny has been placed on providing
justifications for the output of automated
systems\cite{rudin2019stop,mittelstadt2019explaining}. This is
required for a variety of reasons, whether it be regulatory
compliance, ethical analysis, debugging, or further
comprehension of the data used to train the algorithm. For
the purposes of this article, we divide the algorithms in
this field into two categories: transparent (or
intepretable) models and post hoc explanations.  Transparent
models are algorithms that, once trained, can be
intrinsically understood by a particular audience;
examples include shallow decision trees or rule
lists.  Algorithms of this type are generally simpler than
more opaque algorithms and do not include approaches such as
deep neural networks. Post hoc explanations, on the other
hand, are methods that digest and summarize information
from an already optimized external algorithm (such as
feature attribution methods).  While transparent models are
intrinsically interpretable, explainable models are those
which have additional post hoc explanation methods to
provide reduced representations of the knowledge
present in the trained model.

The main contribution of this work is the following
realization: extracting the knowledge present in a
classifier trained to estimate force-field error provides an
explanation of the errors present in the original model. The
nature of the insight provided depends on the particular
methods from XAI we adopt. 
To illustrate this
concept we take a particular modern explanation technique,
Shapley additive explanations (SHAP values), and demonstrate
how they can isolate which physical portions of two CG
proteins (dodecaalanine and actin) are problematic for
specific CG force-fields.  The goal of this paper is not to
model these two proteins accurately; instead, it is to show that
non-ideal force-fields can be detected and understood.
These insights are then aggregated using typical dimensional
reduction techniques to produce collective variables (CVs)
which optimally characterize the types of error exhibited by
a candidate force-field parameterization. These insights are
shown to be useful when considering the behavior of CG
force-fields, and provide a conceptual basis for future
bottom-up error analysis through classification.

\section{Theory}

The techniques presented in this article compare two high
dimensional free energy surfaces, which we refer to as
$\pmf\ppos$ and $\ff\ppos$, where $\pos$ represents 
a sample on the free energy surface. Via the canonical ensemble 
these free energy surfaces define probability densities as
\begin{equation*}
    \pmfdensity\ppos = \pmfpf^{-1} \pmfbfpp
\end{equation*}
and
\begin{equation*}
    \ffdensity\ppos = \ffpf^{-1} \ffbfpp,
\end{equation*}
with $\pmfpf$ and $\ffpf$ defined as the integral of
$\pmfbf$ and $\ffbf$ over all possible $\pos$. These free
energy surfaces could be those defined by any collective
variables. However, in the analysis that follows we will
often assume
that $\pos$ are the configurational variables that comprise
the domain of the configurational part of a CG
Hamiltonian (typically the CG Cartesian
coordinates or pairwise distances); in this case $\ff$
refers to the configurational CG force-field and $\pmf$
refers to the manybody-PMF.

Classification is the machine learning task of predicting
the most probable class, or label, associated with a given
data point\cite{friedman2001elements}.  For example, one might want to predict the
particular number present in a given picture of a handwritten digit.
Algorithms in supervised classification are trained to
complete this labeling task by studying an already labeled
data set: in the previous example, said data set would be a set of
pictures that have had the correct number already associated
with each picture. In certain learning contexts various
samples may not have a clear correct class. For example,
certain hand written digits may be so messy that only the
original writer knows the digit truly intended. In order to
naturally adapt to ambiguous samples, classifiers can be
designed to output a guess of the probability each 
possible
class\cite{friedman2001elements,buja2005loss,gneiting2007strictly,
reid2011}. 
Focusing on the case where we only have two
possible classes to predict (for example, if we were only
considering pictures of 4 or 5), this probabilistic estimate
can be quantified by a single real number for each sample
between 0 and 1.

It is possible to characterize classification tasks
statistically. Again focusing on the task of classifying a picture
as containing a 4 or 5, we first define the distribution of
all possible pictures we could consider (containing 4 
\emph{or} 5)--- this
characterizes the fact that certain images (such as one
comprised of random pixels) are not typical of pictures of
either 4 nor 5. We refer to
this overall probability density as $\marg\ppos$. Then, for each
possible picture, we refer to the conditional probability
that a specific picture contains a 4 as
$\cond\ppos$; the probability of 5 is then $1-\cond$. 
Together, $\marg\ppos$ and $\cond\ppos$ define
a joint probability density over pictures and possible
classes. A classifier that is directed to produce
probabilities attempts to estimate $\cond$ on samples
produced from $\marg$. 

This statistical description can be reframed by considering
the pictures that are characteristic to each class
individually. This involves first considering pictures which
have a ground truth of containing 5, and then those with a
ground truth of 4. The ambiguous pictures above will be then
described as areas in which these two distributions (termed
the class conditional
distributions) overlap. When combined with the overall balance
between the classes,\footnote{The classification
problems constructed in this paper have equal overall 
populations of each label by design, and our expression make
this assumption throughout the article.} the class conditional description in
this paragraph and the picture conditional viewpoint in the
previous paragraph are equivalent specifications of a
classification problem. 

Classification is used in this work by setting the class
conditional probabilities to $\ffdensity$ and $\pmfdensity$.
Using the connections between the two views of
classification, $\cond$ then takes the following form:
\begin{equation}
    \cond\ppos = \frac{\pmfdensity\ppos}{\pmfdensity\ppos +
    \ffdensity\ppos}
\end{equation}
A calibrated classifier can be used to approximate $\cond$
using only samples from each free energy surface;
importantly, this implies that it is not necessary to know
the values of $\ff$ or $\pmf$ evaluated on samples in order 
to estimate $\cond$ on each sample. This approach is
summarized by the following algorithm:
\begin{enumerate}
        \item
            Generate $N$ samples from $\pmf$
        \item
            Generate $N$ samples from $\ff$
        \item
            Label all samples from $\pmf$ with ``A''
        \item
            Label all samples from $\ff$ with ``B''
        \item
            Train a calibrated classifier to predict $\cond$
\end{enumerate}
$\cond$, when combined with $\beta$,  can be transformed
into a offset pointwise difference between $\ff$ and $\pmf$,
which we refer to as $\deltau$:
\begin{equation}\label{eq:deltau}
    \deltau\ppos := \kbt \log \frac{\eta\ppos}{1-\eta\ppos}
    =
    \ff\ppos - \pmf\ppos
    + \kbt \log \frac{\ffpf}{\pmfpf}.
\end{equation}
When $\ff$ corresponds to the configurational distribution
of a CG simulation and $\pmf$ corresponds to the
manybody-PMF, we typically can evaluate $\ff$ on an
arbitrary configuration but are unable to evaluate $\pmf$.
Additively combining $\deltau$ and $\ff$ provides
an estimate of $\pmf$, one that would be exact with a
perfect classifier and learning procedure. In practice,
limited data and imperfect classification algorithms make
this estimate only approximately correct. Using $\deltau$
to form an additive update to $\ff$ has been
performed in the past, see \citet{lemke2017neural}.
This estimation duality,
combined with the realization that classification is a
variational process with respect to the proposed hypothesis,
establish that classification can here be viewed as a
variational technique to estimate the manybody-PMF using
$\ff$ as a reference. Note that $\deltau$ is
defined here such that there is no unknown global offset.

As $\deltau$ precisely characterizes the pointwise difference
between two free energy surfaces, evaluating it at a
particular configuration quantifies the difference in the
conditional free-energies at that point. Areas of high
$\deltau$ imply that one ensemble has considerably more
population in said area than the other ensemble, while a
largely negative $\deltau$ implies the opposite.
Equivalently, when used as a
structural CV, $\deltau$ describes which
configurations occupy areas of high distributional overlap
and which are specific to either free energy surface. In the
context of $\ff$ approximating $\pmf$, linking $\deltau$ to
other intuitive structural variables can characterize what
errors a CG model is committing. For example, if $\deltau$
is negative whenever a particular bond distance is small,
this implies that the CG model ($\ff$) is over stabilizing
small bond distances. The advantage relative to direct
visualization of
configurational statistics from either ensemble is that
configurations occupying areas of high distributional
overlap may be discarded prior to analysis. This approach,
however, is still relatively tedious: it again requires
human analysis of the resulting configurations, incurring all of
the difficulties discussed in the introduction. An appealing
alternative is to understand the algorithmic form of
$\deltau$ itself: for example, if it is a linear function,
its learned coefficients may offer insight. However, if
$\ff$ is composed of low order $n$-body terms while $\pmf$
contains higher order terms, $\deltau$ will contain high
order terms as well, and it may be difficult for
intepretable models to provide a good estimate of
$\deltau$.  In
this article, we use techniques from XAI to overcome this
difficulty and extract configurational information from a 
complex estimate of $\deltau$. 

\subsection{Feature attribution}

XAI includes of a large variety of methods. This article
will focus on the use of a single method from this field:
SHAP values\cite{lundberg2017unified,
lundberg2020local2global,lundberg2018explainable}. SHAP values 
are a feature attribution method\cite{molnar2020book,
kodratoff1994,ruping2006learning}
with a strong mathematical underpinning. Feature attribution
methods or feature importance measures provide a
quantification of how informative a feature, or particular
input variable, is to an algorithm. Some feature attribution
methods are global, meaning that stated feature values are
related to the aggregate behavior of the classifier over the
entire data set. Other feature attribution methods, such as
SHAP values, are local: every prediction made by a
classifier can be associated with a particular set of
SHAP values.  When estimating $\deltau$, this means
that a prediction of a large positive or negative $\deltau$ can be
analyzed to determine which features lead the classifier to
that conclusion.\footnote{Here, $\cond$ is not approximated as
solely the output of the gradient boosted trees; it is the
composition of said output with the distance matrix
featurization. As a result, feature attributions information
is produced on the resolution of the distance matrix, not the Cartesian
coordinates.}

The classification examples in this article quantify the
error present in various CG models of proteins. The
classification algorithm is trained on the ordered distance
matrix derived from each configuration. As a result,
applying a feature attribution method to explain $\deltau$
isolates which pairwise distances are most connected to the
estimated error, and in doing so clarifies which
configurational aspects of the protein are well captured by
the CG model.

\subsection{Shapley and SHAP values}

SHAP values are based on Shapley
values\cite{shapley1953value,molnar2020book} from cooperative
game theory. We first explain Shapley values and how they
relate to classification, and then provide a description of
SHAP values.

\subsubsection{Shapley values}

Shapley values are a method to fairly distribute a reward
among a group of individuals. Suppose a group of five
scientists decides to create a product to bring to market.
These five people have joined together because their
individual knowledge, when combined, produces a better
product then they could individually. However, suppose one
of the five people has knowledge that is vital to the
product: if they were not present, the total amount of
profit would be greatly diminished. In contrast, the
expertise of the remaining four people is largely, but not
completely, shared. As a result, losing one of those four
people would reduce the possible profit, but would not do so
substantially. In this situation, how should the profit be
fairly divided among the scientists? Allocation in these
circumstances is answered by Shapley values.

The central calculation needed to define Shapley values is
the ability to estimate the reward had some of the
individuals in the group not been present.  Suppose the five
people present are referred to as $A$, $B$, $C$, $D$, and
$E$. We then denote the reward when everyone is present
$\cost(\{A,B,C,D,E\})$.  In order to calculate Shapley values we
must be able to calculate, as an example, $\cost(\{A,B,D,E\})$:
the reward had individual $C$ not been present. Shapley
values then consider growing the number of present individuals
incrementally, such as the (ordered) sequence
$\cost(\{B\}),\cost(\{B,A\}),\cost(\{B,A,D\})...$, and associating with
individual $D$ the term $\cost(\{B,A,D\})-\cost(\{B,A\})$: the incremental
increase that was seen when $D$ was added in this particular
sequence. The Shapley value averages over all such
sequences.
Mathematically, the Shapley value for player $i$ is defined as
follows,
\begin{equation}\label{eqn:shapley}
    \phi_i = 
    \frac{1}{n!} 
    \sum_{R} \cost(P^R_i \cup \{i\}) - \cost(P^R_i) 
\end{equation}
where $n$ is the number of individuals, $R$ iterates over
all possible \emph{orders} (not subsets) of players and
$P^R_i$ is the subset of individuals that precedes player
$i$ in that particular $R$. It is important to note here
that this sum is over all possible orderings, as where
$\cost$ only depends on the members present, not the order
in which they were added. This calculation must be
performed for each individual (or player) for which we wish to
calculate a Shapley value. 

Shapley values satisfy a number of intuitive mathematical
properties\cite{shapley1953value,young1985monotonic,
lundberg2017unified} and in some cases are the only allocation method
that does so. The most important property for the current
application is that summing together the Shapley values for
all players provides the original reward when the entire
group is present.

\subsubsection{SHAP values}

The connection between Shapley values and feature
attribution is made by considering every individual prediction made by
a classifier a game in which each feature is a player. The
output of the game, in analogy with the total profit in the
previous subsection, is the numerical prediction of the
classifier. However, one important detail is absent when
considering feature attribution: what does it mean for a
feature to be \emph{missing}? It is possible in some cases
to retrain a model with only a subset of the original
features\cite{sundararajan2019many}; however, the number of models required quickly
becomes infeasible. Instead, Shapley Additive Explanation
(SHAP) values train a single model and average said full
model's predictions over the missing variables to represent
missing values\cite{lundberg2017unified}. For example,
consider hypothetical classifier $f(w,x,y,z)$ with four
input variables. Suppose we wished to calculate the Shapley
value of $w$ for configuration $(w_0,x_0,y_0,z_0)$: this
would include calculating $f_{wz}(x_0,y_0)$, where $w$ and
$z$ are ``missing''. SHAP values dictate that
$f_{wz}(x_0,y_0) := \int f(w,x_0,y_0,z) p(w,z|x_0,y_0) dwdz
$, where $p(w,z|x_0,y_0)$ is the distribution of $w$ and $z$
\emph{conditional}\footnote{Some
implementations of SHAP values instead use the marginal
expectation value, some label the 
choice of the marginal
expectation as an approximation,
and some argue the marginal value is the
correct one to use from an interventional 
perspective\cite{lundberg2017unified,sundararajan2019many,
kumar2020problems}. The
implementation and physical interpretation presented in this 
paper use the conditional
expectation value.} on $x=x_0$ and $y=y_0$. With this definition of
``missing values'', Shapley values are applied as before to
give SHAP values for each feature. The algorithmic
calculation of these values for arbitrary classification
techniques is far from trivial, but is possible
for tree ensembles such as those used in this
article\cite{lundberg2020local2global}.

Despite the abstract description above, SHAP values can be
physically interpreted in the current study. All the
examples in this paper perform calibrated classification
using distance matrices as input. The signal additively
explained by the given SHAP values corresponds to $\deltau$.
The individual terms in the Eq. \eqref{eqn:shapley} can be
understood as follows: $\cost(P^R_i \cup \{i\})$ corresponds
to the mean $\deltau$ found when holding the distances
specified by $P^R_i \cup \{i\}$ constant and letting the
remainder of the protein freely explore the canonical
ensemble, and $\cost(P^R_i)$ is the same but letting
distance $i$ also freely explore\footnote{SHAP values 
include a single global offset in their additive
explanations;
however, this distinction does not matter for applications
presented in this article}. In this way, the SHAP
values isolate which parts of a specific protein configuration
contribute to its specified $\deltau$. The same idea can be applied to an
feature set of an arbitrary free energy surface:
the system is allowed to explore conditional to the given
features under investigation.

SHAP values provide a real number for each feature and
configuration considered.  In this way they can be
considered a new set of descriptors for each protein
configuration.  This set of descriptors is of equal
dimensionality to the original data set, and at first seems
to provide little advantage relative to the original
coordinates. However, three properties of SHAP coordinates
are more desirable than the original coordinate system:
\begin{enumerate}
    \item
        SHAP values monotonically relate to $\deltau$.
    \item
        SHAP coordinates are of the same scale for each
        configuration and feature. As a result, the relative
        importance of an inter-domain distance and an
        intra-domain distance can be directly compared in
        SHAP space.
    \item
        SHAP values individually reflect manybody correlations in the
        original data.
\end{enumerate}
As a result, quickly inspecting the individual ranges of SHAP values
(using, for example, box plots) can produce insight
about problematic aspects of $\ff$ in the original
coordinate system. Additionally,
we reduce the dimensionality of the generated SHAP coordinates to produce a
lower dimensional set of CVs that summarizes the types and
severities of errors present in a given CG
model. These analyses suggest aspects of the
CG model's force-field basis that may be
modified to improve accuracy. It is important to note
that SHAP values using conditional expectations, such as
those used in this article, assign feature attributions
using both correlations present in the learning
distribution (here, the combined model and reference
ensembles) combined with information in
$\deltau$\cite{sundararajan2019many,kumar2020problems}. For
example, consider the more generic 
case of the function
$f(x,y) = 2x$ analyzed over a distribution of $x$ and $y$ where
$x$ and $y$ are highly correlated. Conditional SHAP values,
such as those in this paper, will assign importance to both
$x$ and $y$: this is informally because knowing that $y$
is large also implies that $x$ is large due to the
correlation: $y$ contains information about $x$. As a
result, it is difficult to infer the structure of a
hypothetical $f$ using said SHAP values. Other feature
attribution methods based on Shapley values circumvent this
limitation at the cost of considering unrepresentative
samples\cite{sundararajan2019many}; we leave investigating 
these alternative methods to a future study.

\section{Methods} \label{methods}

The methods used in this article can be reproduced using
publicly available libraries.  Classification was performed
using the DART (boosted decision trees with
dropout) algorithm in the Light Gradient Boosting
Machine (lightgbm) library.  This library supports
a large number of hyperparameters.  Unless otherwise
specified, hyperparameters were set to 5 leaves, 5000 trees,
and 6150 bins per feature; trees were dropped (DART) at a rate
of 0.5 with a max drop of 1000, and each learned tree saw
80\% of the feature space and 15\% of the full data set
(this random subset of data was redrawn every 50 trees).
50\% of each molecular data set was used as a training set,
while the other 50\% were used as a test set.  Trees were
grown to minimize the log loss. Subsampling
data and features sped up training.  The results presented
are obtained from the test set; however, due to the
regularization imposed, performance on the train and test
sets were close to identical.  Qualitative conclusions made
through CVs generated via dimensional
reduction were stable to choices of hyperparameters.
Estimated pointwise free energy differences were found to be
somewhat sensitive to choices of hyperparameters. This is
expected: large absolute differences in $\deltau$ imply that
a comparison is being made at a location with very low
configurational density in one ensemble; the precise level
of population is difficult to accurately estimate without
using enhanced sampling.
As a result, if one wishes to make a quantitative comparison
between models based on the pointwise free energy
differences, a large amount of well sampled data must be put
through a very careful train-test-validation based
framework. In contrast, the approach in this paper used
classifiers with similar levels of regularization
throughout.

Dimensional reduction was performed using principal
component analysis (PCA) combined with the Uniform Manifold
Approximation and Projection (UMAP) method\cite{mcinnes2018umap}.  
This technique
was selected due to observed computational efficiency and
separation of clusters, not due to any physical argument.
Results generated using t stochastic neighbor
embedding\cite{maaten2008visualizing}
(t-SNE) appeared different, but lead
to similar physical conclusions.  It
was observed that depending on the system under study, the
structure of the high dimensional data varies greatly (for
example, in terms of connectivity), suggesting that
different models in the future may require different analysis strategies.
Unless otherwise noted, 20 PCA coordinates were used as
input to UMAP to create a 2 dimensional projection using 64
nearest neighbors (other parameters were left to their
defaults).  Qualitative aspects of the generated CVs were
found to be consistent across a wide range of parameters;
said high number of neighbors was used to avoid small clumps
of points which did not improve physical interpretation.
$k$-nearest neighbor regression (KNN regression) was used to
map coordinates to UMAP coordinates using the following
procedure. First, data was mapped to SHAP values; second,
PCA was used to map the SHAP values onto the first 20
principal modes. UMAP was then applied to map these PCA
coordinates to UMAP space. This data served as a training
set for KNN regression, which was parameterized to map SHAP
values to UMAP coordinates. New configurations were
processed by first producing their SHAP values; these SHAP
values transformed using the trained KNN regressor. When
projecting data onto SHAP coordinates which were generated
from other ensembles, KNN regression was always used. For
dodecaalanine, KNN
regression was trained using 2$e$4 data points; the
remainder of the samples in each ensemble
were also projected using KNN regression. In the case of actin,
all 1$e$4 samples were processed directly with PCA and UMAP.
KNN regression was parameterized to use the 5 closest
neighbors.

\subsection{Dodecaalanine}\label{ddamethods}

Dodecaalanine (DDA) is a short polypeptide which adopts a
variety of conformations in solution: a hairpin like
conformation, a helical conformation, and an extended
conformation (see, for example, \citet{rudzinski2015bottom}).  
DDA was simulated 
at the atomistic resolution
using Amber18\cite{amber18} and the
Charmm36m\cite{huang2017charmm36m} force-field.  Each of two
replicas was as solvated with 7121 water molecules (TIP3P)
and 20 sodium and chloride ion pairs. This system was
relaxed using steepest descent for 5000 steps, followed by
equilibration via NVT simulations and NPT simulations.
Production samples were extracted every 50 $ps$ from a 5.1 $\mu s$
trajectory propagated using a 2 $fs$ timestep in the NVT
ensemble at 303$K$ using a
Langevin\cite{schneider1978molecular} thermostat with a
damping parameter of 0.5 $ps$. Hydrogen bonds were
constrained via SHAKE.  The resulting DDA frames were mapped
to a resolution of one CG site per amino acid; coordinates
were determined by a center of mass mapping. CG simulations
were propagated in LAMMPS\cite{plimpton1995fast} (version
lammps-7Aug19, lammps.sandia.gov) and were
started from an initial structure obtained from the mapped
atomistic ensemble. The system was minimized for 5$e$5 NVE/limit
steps, then propagated for 5$e$5 steps using a
Nose-Hoover\cite{shinoda2004rapid}
thermostat with a 0.2 $fs$ timestep and a coupling
parameter of 0.5 $fs$.  Production CG simulations were propagated
using using a Langevin thermostat with a 2 $fs$ timestep and
a damping parameter of 0.1 $ps$. 1.8$e$5 total CG samples
were taken by sampling every 0.4 $ps$.  Each DDA
classification task was performed on 3.6$e$5 samples: 1.8$e$5
were randomly sampled from the mapped atomistic replicas,
and 1.8$e$5 were taken from a contiguous CG trajectory. 

\begin{figure}
    \centering
    \centerline{\includegraphics[scale=0.18]{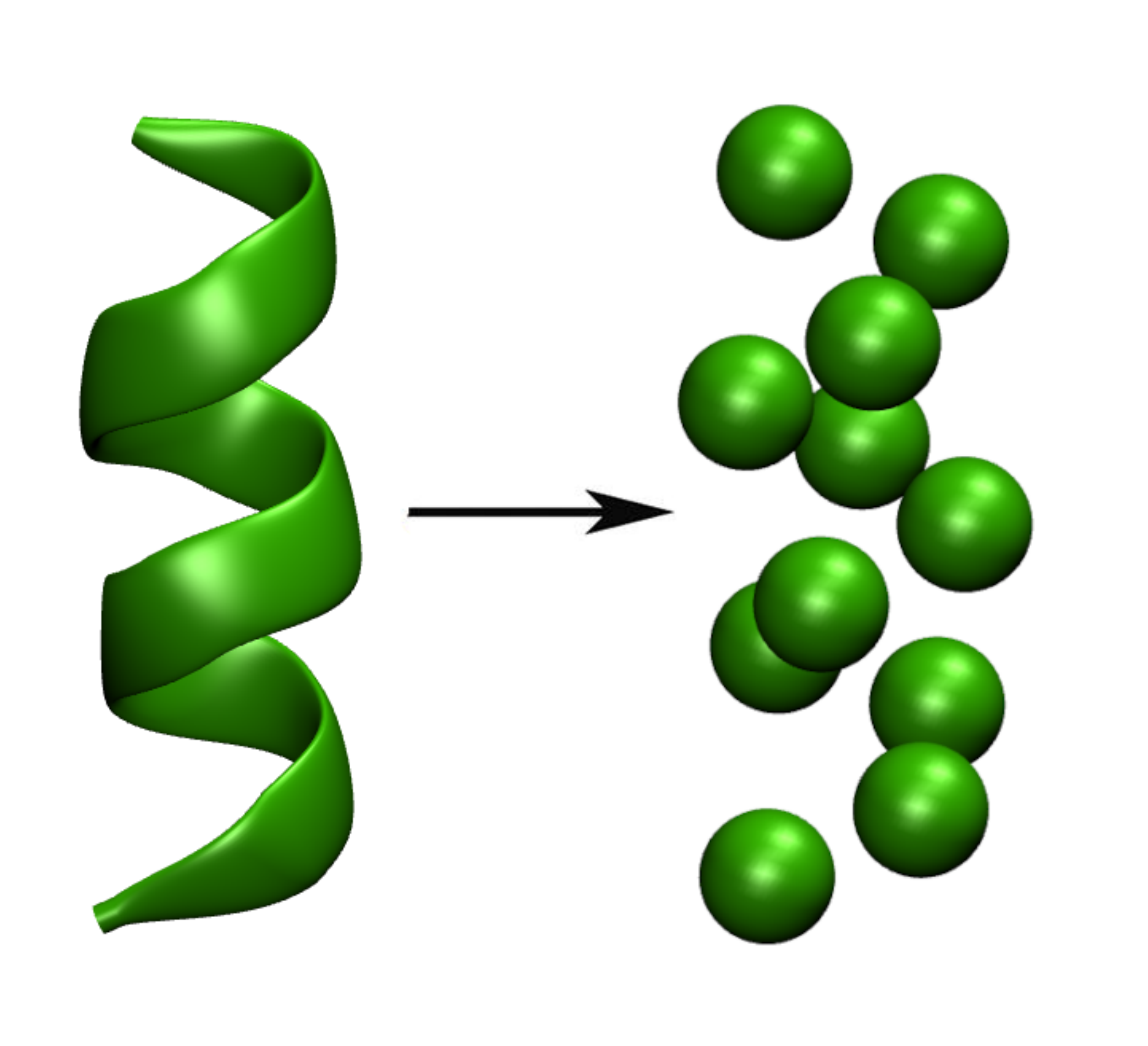}}
    \caption{The map used to coarse-grain dodecaalanine in the
    current study. Each amino acid was mapped to a single CG
    site via a center of mass mapping.}
    \label{fig:ddamap}
\end{figure}
Relative entropy minimized (REM) models of DDA were
iteratively parameterized using the Adversarial Residual
Coarse-Graining\cite{arcg} software framework
(git repository available upon request). Analysis and
visualization used the theano\cite{theano}, 
numpy\cite{numpy}, lightgbm\cite{lightgbm},
scikit-learn\cite{scikit-learn},
shap\cite{lundberg2017unified,treeexplainer}, 
pandas\cite{pandas113},
ggplot2\cite{ggplot}, data.table\cite{datatable}, and 
pracma\cite{pracma} libraries
Each iteration's CG simulation was performed using
the settings described above for CG models. Marginal
distributions were confirmed to well approximate those
implied by the theory underpinning REM (SI). All potentials
(bonded and nonbonded) were represented via pairwise
interactions. Nonbonded pairwise potentials
were represented using a WCA potential summed with a 3rd
order $b$-spline (the additive non-spline terms significantly improve
stability during training); bonded potentials were represented using a
harmonic potential summed with a 3rd order $b$-spline. For
nonbonded interactions, 20 equally space spline knots were used for both
the bonded and nonbonded splines; the nonbonded splines and
tables ranged from 0.001 to 20 Angstroms, and the bonded splines and
tables ranged from 0.001 to 10 Angstroms. Both the
parameters controlling the WCA and harmonic terms were
transformed through $\exp$ to enforce positivity. Bonds were
present both between nearest neighbors along the CG backbone
and neighbors separated by a single site; in the latter
case, the bonded interactions emulated angle potentials. At each 
iteration, 1.8$e$5 configurations were
randomly sampled from atomistic trajectory and used to
calculate the REM derivative (along with the entire CG
trajectory). Larger sets of configurations for derivatives
were not seen to change the REM results, although smaller
sets (e.g., 9$e$4) were noted to give biased
potentials. For each iterative update, the next set of parameters was generated using
the RMSprop algorithm\cite{hinton2012neural} using a rate of 0.009 for all
parameters (all parameterization was performed relative to
LAMMPS real units). Resulting parameter updates were clipped
to a absolute maximum of 0.07. Parameters were updated until
no reliable change in any potential was visible for 30
iterations.

The radius of gyration and Q-helicity were used to quantify
the behavior of DDA; their formulation can be found in
\citet{rudzinski2015bottom}. Q-helicity quantifies the
similarity of a given configuration to a helix:
0 corresponds to no helical character, while 1 corresponds to
a completely helical polypeptide. Similar CVs are present in
enhanced sampling studies, see \citet{piana2007bias} and
\citet{prakash2018peptoid}.

\subsection{Actin} \label{actinmethods}
\begin{figure}
    \centering
    \centerline{\includegraphics[scale=0.9]{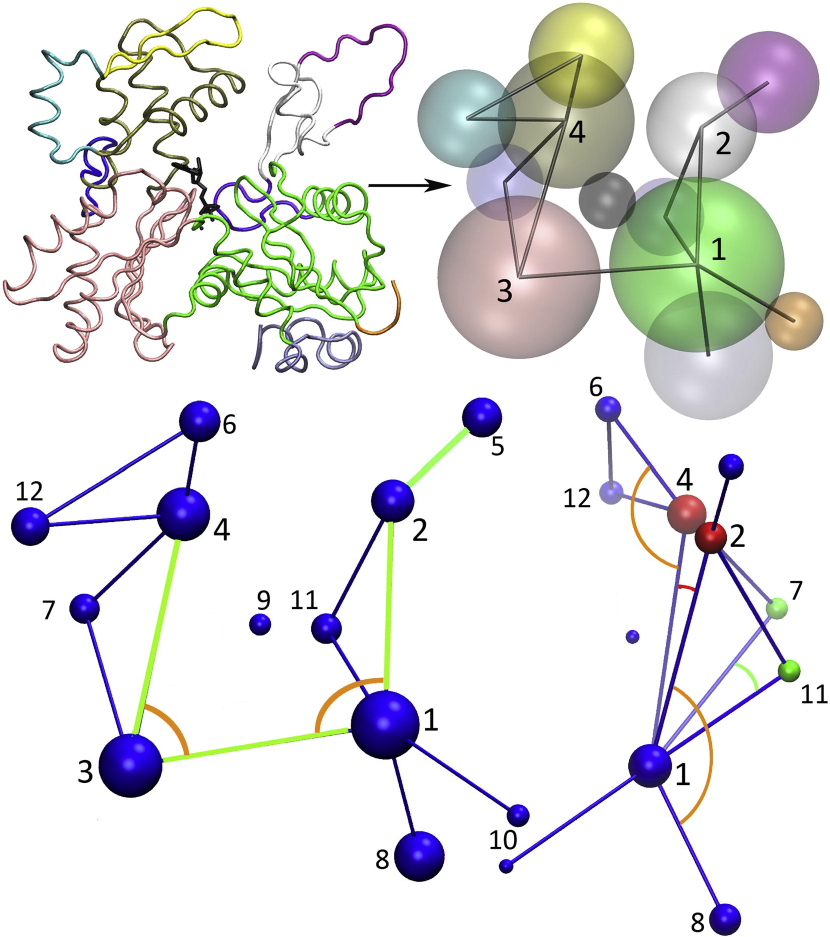}}
    \caption{The map used to coarse-grain actin in the
    current study. Each set of atoms was mapped to a
    position via a center of mass mapping. CG models 
    of actin were simulated and modeled at both the
    resolution of all 12 sites presented and at the
    resolution of sites 1-4.}
    \label{fig:actinmap}
\end{figure}

Actin was simulated at the atomistic resolution using
GROMACS 5.1.4\cite{gromacs5}; the protein was solvated with 29530 waters 
(TIP3P) and 107 pairs of potassium
and chloride ions.
Equilibration details may be found in
\citet{hocky2016cations}. Production
simulations were performed for 1 $\mu s$ using
CHARMM27+CMAP\cite{mackerell2004extending}
in the NPT ensemble at 310$K$ and 1 $bar$ using the
stochastic velocity rescaling thermostat\cite{bussi2007canonical} 
with a coupling
parameter of 0.1 $ps$ and a Parrinello-Rahman
barostat\cite{parrinello1981polymorphic} with
a coupling parameter of 2 $ps$. Hydrogen bonds were
constrained via LINCS. Samples were
collected every 100 $ps$. The atomistic frames were mapped
to the CG resolution using the map found in
\citet{saunders2012comparison}. Briefly,
sites indexed 1-4 represent actin's four main subdomains
which are approximately arranged at the 4 corners of a
square; site 9 represents the nucleotide ADP situated at the center
of these 4 subdomains, and site 5 represents the D-loop, a
semistructured region connected to site 2. The map is
characterized in Fig. \ref{fig:actinmap} (adapted from 
\citet{saunders2012comparison} with permission). Hetero-elastic network models (hENMs) 
were created using the procedure in \citet{lyman2008systematic}
using this atomistic 
trajectory. In certain portions of the results, models are
parameterized using all 12 of these sites. In other portions
only sites 1-4 are used.
CG samples of actin were generated using LAMMPS (version
17Nov2016). Initial configurations were taken from the
mapped atomistic trajectory. The system was propagated using
a timestep of 1 $fs$ via a Langevin thermostat with a
damping parameter of 50 $fs$. 2$e$5 CG samples were obtained
by sampling every 10 $ps$. The relaxation steps used for
actin were less aggressive than those used in DDA as no REM
optimization was needed. When performing classification
based comparisons, similar procedures were followed except
the size of the data was significantly smaller: 1$e$4
samples were used from each data source (the entirety of the
mapped atomistic simulation was used, and the CG trajectory
was randomly subsampled).

\section{Results}

SHAP coordinates were used to explain errors in models of
DDA and actin. As elaborated on in section \ref{ddamethods}, when 
considering DDA a single resolution of
12 sites was used with multiple pairwise spline
force-fields, each of which was minimized using REM.
The CG resolution was
defined as the center of mass of each amino acid. 
In the case of actin, models
consisting of fully connected networks of harmonic springs
at two resolutions (12 and 4 sites) were studied, each of which were
parameterized using the hENM method as described in section
\ref{actinmethods}; as a result,
all harmonic springs present differed. 
Section
\ref{ddaresults} studies the effect of various force-field bases on DDA,
providing examples of using SHAP based analysis to
understand and improve existing force-fields. Section
\ref{actinresults}
analyzes the role of the CG map when modeling actin,
providing an example of how $\deltau$ and the resulting SHAP
coordinates relate to selecting  model resolution.

\subsection{Dodecaalanine}\label{ddaresults}

Dodecaalanine was modeled using five different force-fields.
Each force-field was composed solely of pairwise
interactions at the bonded and nonbonded level. Sites
adjacent along the backbone were connected via bonds; sites
separated by a single site were connected via an additional
bond to emulate a angle potential. Within each type of
interactions (bonded, angle-bonded, and nonbonded), a unique
interaction type was defined for each possible pair of
site types. Each  model can be considered as extending the model
before it.

The first model was composed of a single site type. The
pairwise nonbonded interactions were set to be constant,
i.e. nonbonded pairwise forces were uniformly zero. This
resulted in unique two pairwise interactions (one bonded and
one angle bonded).  The second model was identical to the
first model, but included pairwise nonbonded interactions.
The third model then included a single additional collective
site type for the termini, resulting in two site types
total. The fourth extended the model by considering the site
types at each termini to be distinct, resulting in three
site types.  The final model was extended to five site types
by providing additional unique site types to each CG bead
adjacent to a termini bead. The difference in the pairwise
potentials derived when adding new site types is
substantial, as shown in the SI. Only a subset of these models is
analyzed in certain sections for brevity.

\begin{figure}
    \centering
    \centerline{\includegraphics[scale=0.6]{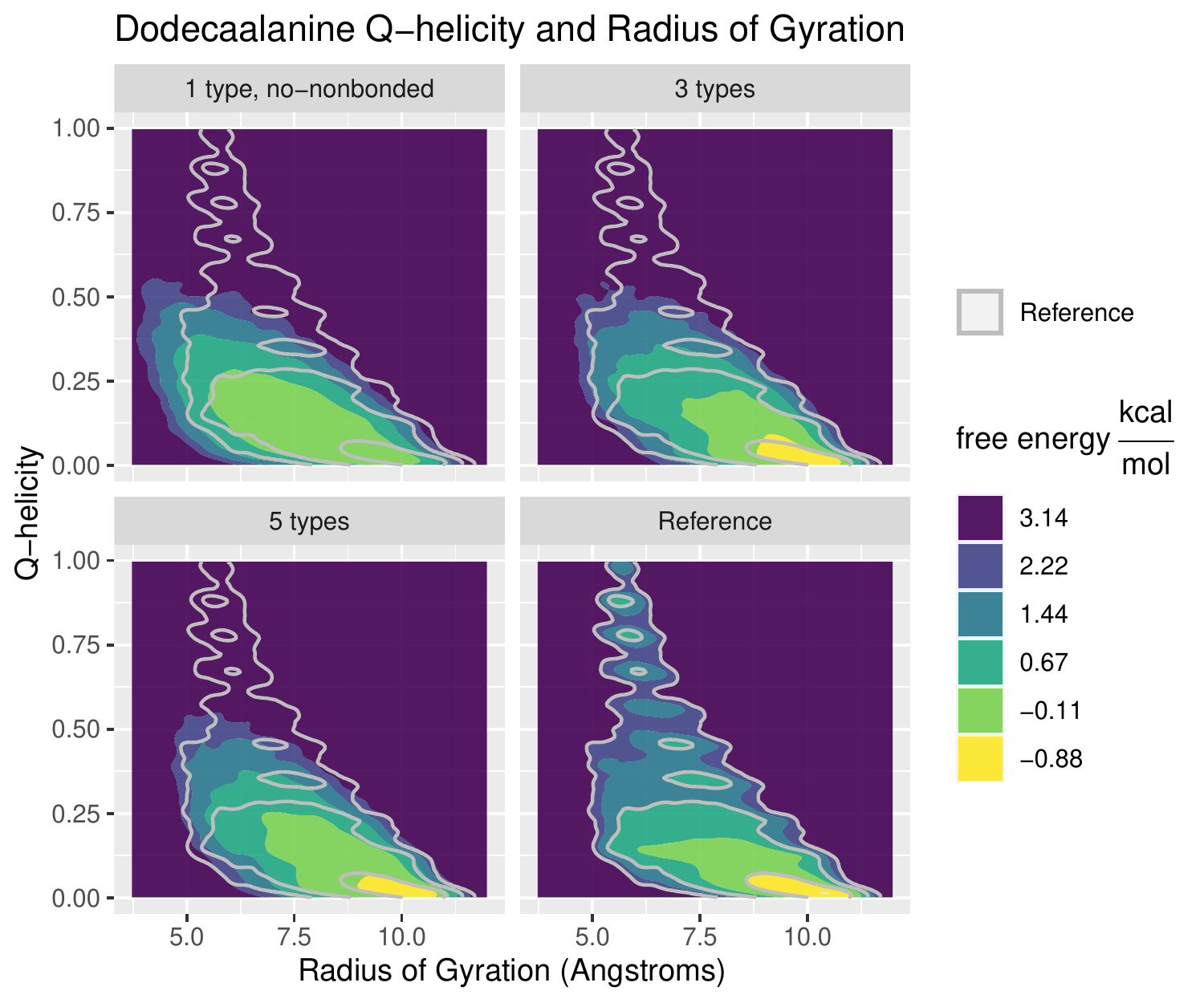}}
    \caption{Free energy surfaces of the
    radius of gyration and Q-helicity for various models and
    the reference distribution. The grey overlay (and lower
    right panel) is given by
    the reference density and is present as a visual guide. 
    Clear differences exist
    between the models and the reference distribution, but
    little difference can be seen between various models,
    despite significant differences in pairwise potentials
    (see SI).}
    \label{fig:rogqhel}
\end{figure}

Established CVs, 
such as the radius of
gyration and Q-helicity (see section \ref{ddamethods}), show 
significant differences
between the reference atomistic data and the various models:
high helical states are not captured. The difference between
a select few models and the reference free energy surface is
shown in Fig. \ref{fig:rogqhel}.  However, little difference
is apparent between various CG force-fields, save for a
small amount of extra distortion present in the single site
model lacking nonbonded interactions.

\begin{figure}
    \centering
    \centerline{\includegraphics[scale=0.6]{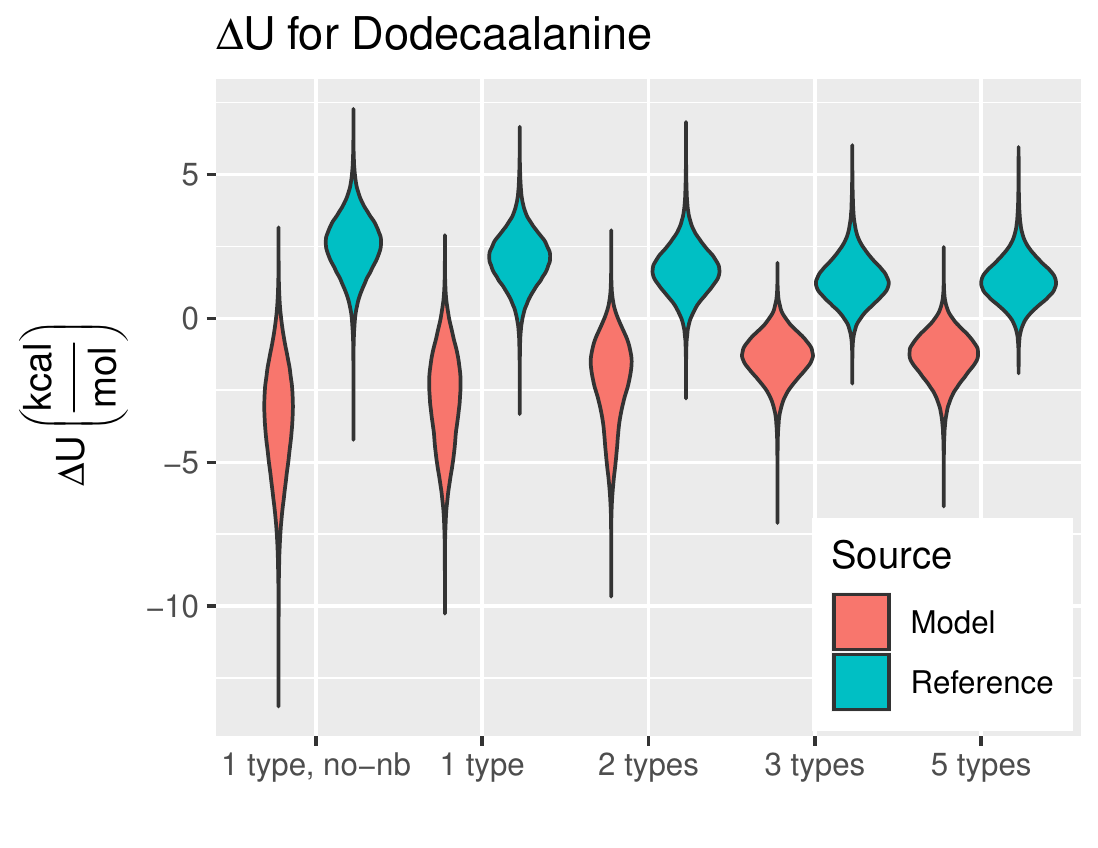}}
    \caption{Violin plots of 
    $\deltau$ for multiple models of dodecaalanine. Each
    model distribution 
    is divided along the reference and
    model ensembles. Note that varying models modify
    $\deltau$, changing the shape of the projected reference 
    distribution.}
    \label{fig:ddaviolinerror}
\end{figure}

Classification was performed between each model and the
reference data, with the results shown in Fig
\ref{fig:ddaviolinerror}. We note that
Fig. \ref{fig:ddaviolinerror} characterizes the distribution of errors at the
resolution of the CG Hamiltonian, but that $\deltau$
is also a valid structural CV\cite{sultan2018automated} that is optimized to
separate the pair of ensembles it is generated from. $\deltau$ additionally 
numerically corresponds to
the difference in pointwise free energy values \emph{at the
resolution given by considering $\deltau$ as a CV}\footnote{If not true, 
there exists a $\cond_{\deltau}$ at the
resolution of $\deltau$ with
less error, which implies that the pullback of $\cond_{\deltau}$ would 
also
outperform $\cond$ at the model resolution, which is a
contradiction if using a proper loss function. This 
is implicity used when featurizing
classification problems. Equality of $\cond$ implies
equality of the relative entropy at the model and  $\deltau$
resolutions.}. There exist other CVs that 
equivalently recapitulate $\deltau$, including the distribution in the full
SHAP feature space. The dimensionally reduced coordinates
are not guaranteed to share this property, but seem to recapitulate
the overlap shown by $\deltau$ alone.

In contrast to the lack of change present in Fig.
\ref{fig:rogqhel}, increasingly complex force-field bases
produces notable changes in the distribution of $\deltau$.
More significant drops in the magnitude of error can be seen after
addition of nonbonded interactions and the introduction of a
third site type. Interpreting $\deltau$ as a typical CV, it
is also apparent that all of the models considered have
relatively poor overlap with the reference data set.

\subsubsection{Uniform no-nonbonded model}
\begin{figure}
    \centering
    \centerline{\includegraphics[scale=0.6]{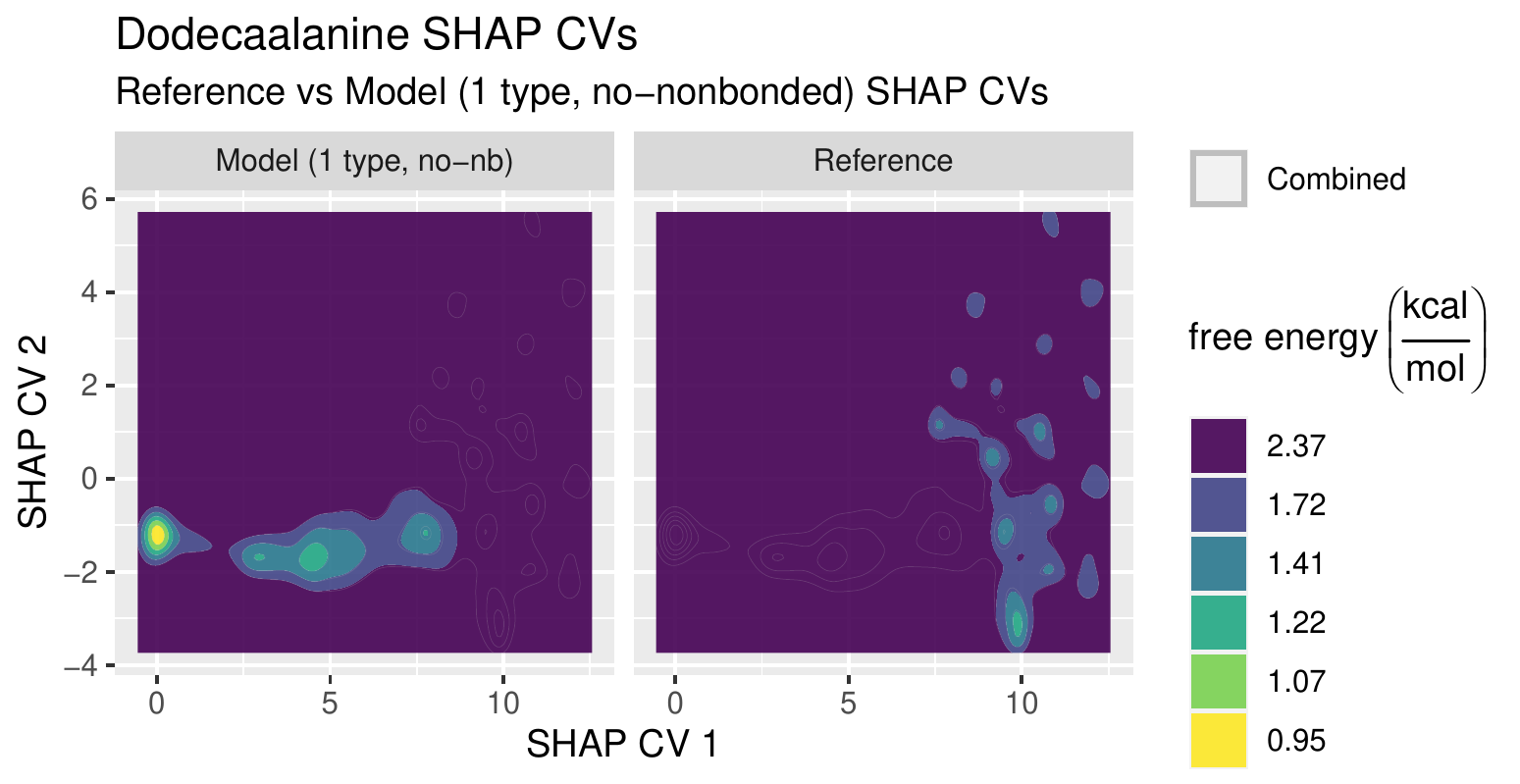}}
    \caption{Free energy surfaces produced along the
    SHAP CVs generated from the 1 type no-nonbonded
    model. Light grey lines characterize the combine density
    of the model and reference data and serve as a visual
    guide.}
    \label{fig:nonb}
\end{figure}

\begin{figure}
    \centering
    \centerline{\includegraphics[scale=0.6]{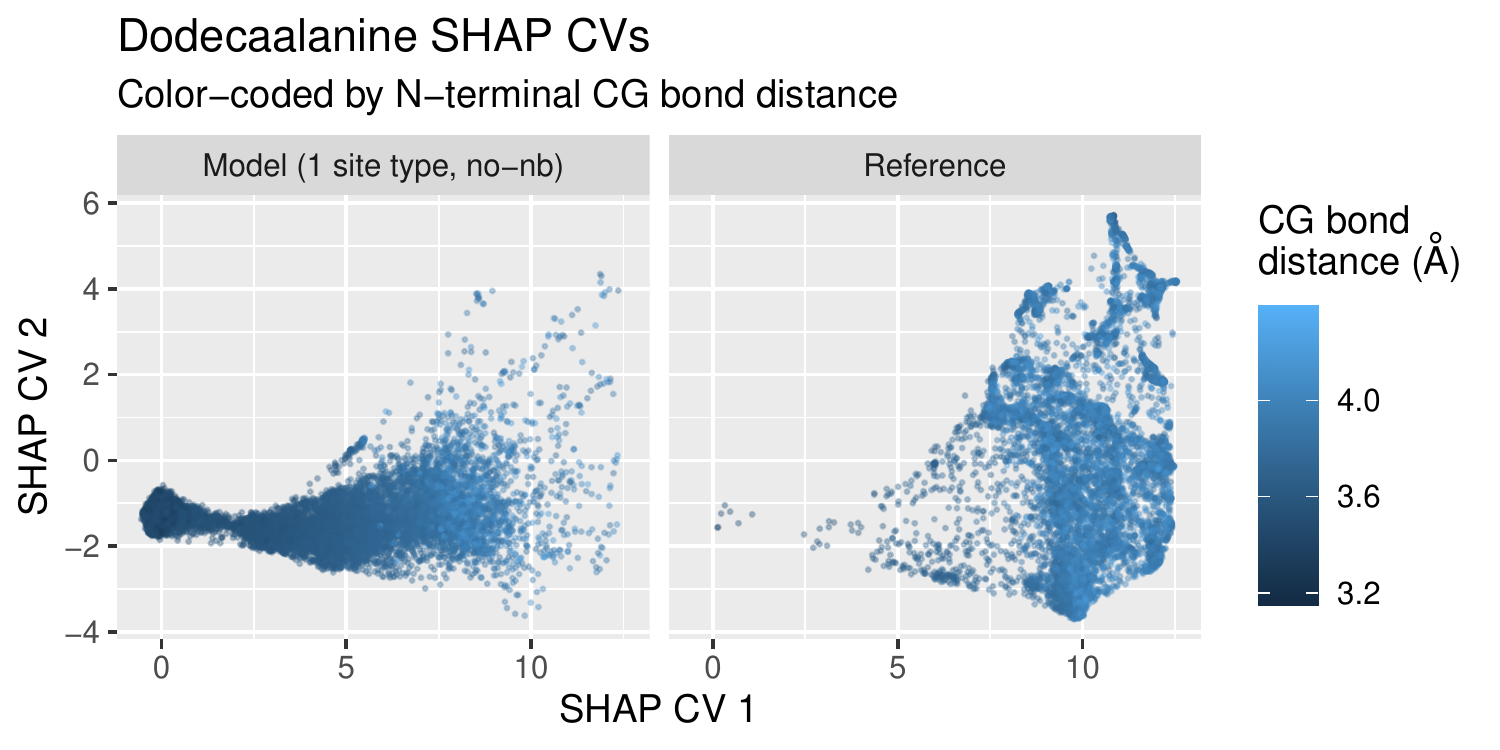}}
    \caption{Samples from the 1 type no-nonbonded and
    reference ensembles projected onto the SHAP CVs in Fig.
    \ref{fig:nonb}. Samples are colored by the distance
    between the N-terminal CG site and its nearest bonded
    neighbor.}
    \label{fig:nonbcolor}
\end{figure}

SHAP variables were constructed for the 1 type DDA model lacking
nonbonded interactions 
using the per-feature output of TreeExplainer as input for
UMAP based dimensional reduction (Fig. \ref{fig:nonb}). 
The lack of overlap seen in Fig. \ref{fig:ddaviolinerror} is
preserved, along with additional organization representing
the various sources of error. While UMAP
lowers dimensionality to allow visualization,
it does not provide explicit formulas for the resulting CVs. 
However, examination of the individual aggregated
SHAP values themselves provides immediate insight into the distances
characterizing $\deltau$. For example, the feature with the largest mean
absolute SHAP value by a factor of 3.5 is the N-terminal CG
bond; the next set of maximal SHAP values spans distances
between sites separate by a single site (e.g.  
backbone index 3 and backbone index 5). Plotting the
N-terminal bond distance with the SHAP CVs (Fig.
\ref{fig:nonbcolor}) demonstrates that the elongated error
present from $(0,-1.7)$ to $(8,-1.7)$ is primarily due to
this bond distance. Larger scale N-terminal effects
(combined with minor distortion at the C-terminus) dictate
the width of this feature. Comparison of SHAP values
via the generated SHAP CVs provides information on their
interdependence. For example, comparison of the values of
radius of gyration and Q-helicity with the generated SHAP
coordinates (SI) implies that the variety of small islands
present in the right of the reference ensemble primarily
represent both partially helical states and minor bond
distance errors in the middle of the protein, while the
fully extended configurations are diffusely present around
$(8,-1.5)$. While the helical errors are spread across
multiple features, they correspond to single islands in the
SHAP based projection.

\begin{figure}
    \centering
    \centerline{\includegraphics[scale=0.6]{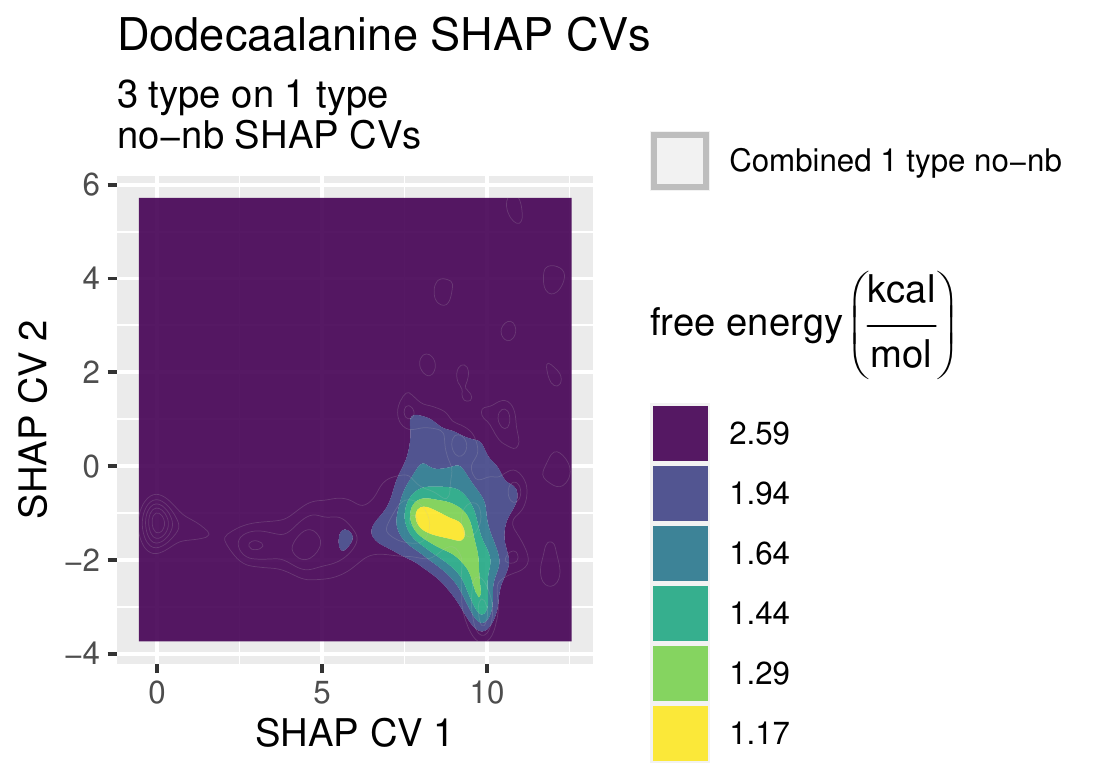}}
    \caption{3 type free energy surface projected along the
    SHAP CVs generated from the 1 type no-nonbonded
    model. Filled surfaces represent the 3 type model
    ensemble, while grey lines represent the combined 1 type
    no-nonbonded and reference ensembles presented in Fig. 
    \ref{fig:nonb}.}
    \label{fig:nonbproj}
\end{figure}

Both non-bonded interactions and a single
additional type for the termini (the 2 type model) each
reduce the elongated error feature dominating the model
lacking nonbonded interactions; however incorporating a
distinct bead for each each termini (the 3 type model) fully
removes the elongated error present in Fig. \ref{fig:nonb}, as 
can be seen by projecting
the 3 type model ensemble onto the coordinates 
present in \ref{fig:nonb} (Fig.
\ref{fig:nonbproj}). The
additional site types also slightly increase the density
present in various partially helical states, but not
sufficiently to register in Fig. \ref{fig:nonbproj}.

\subsubsection{Three-site model}
\begin{figure}
    \centering
    \centerline{\includegraphics[scale=0.6]{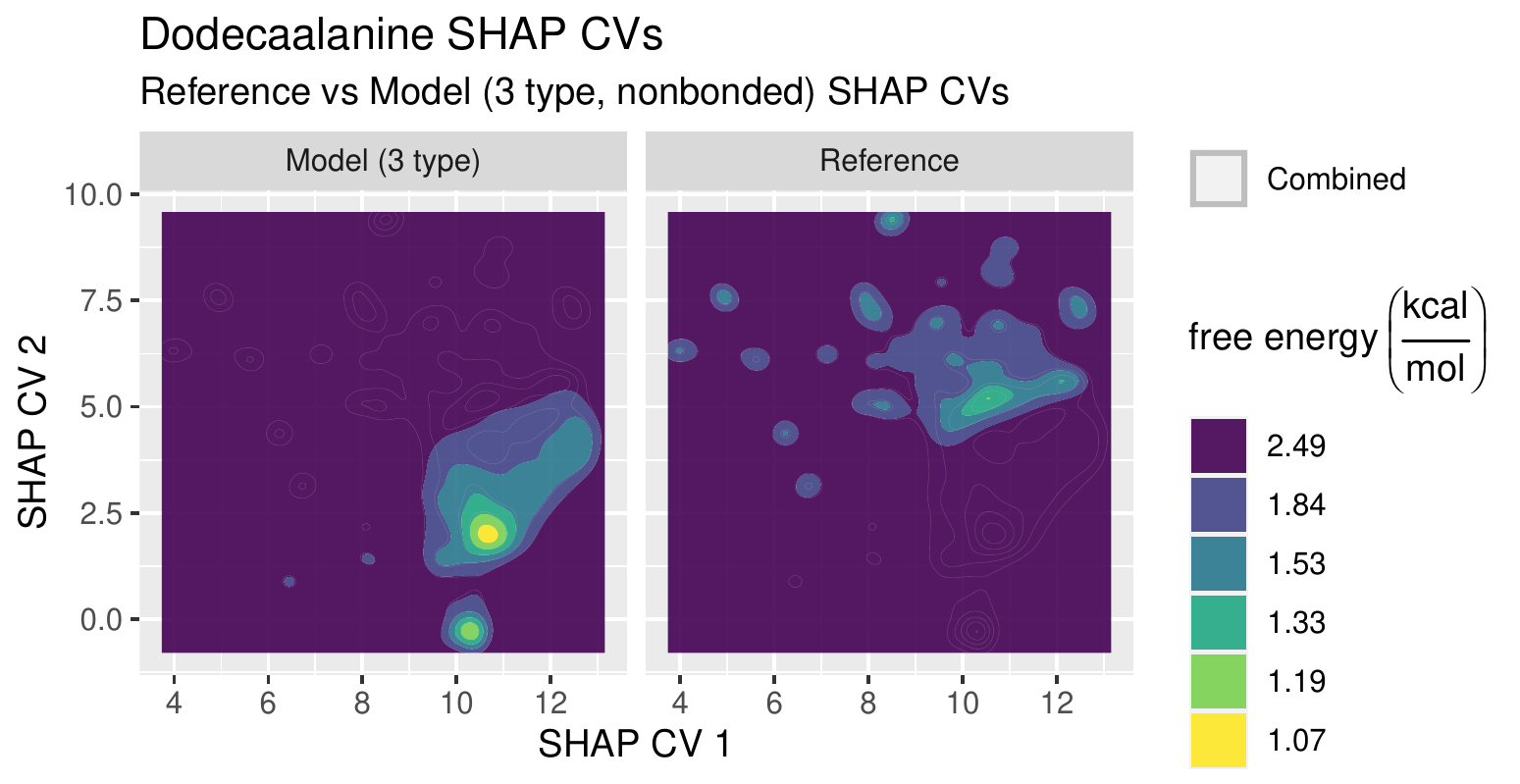}}
    \caption{The free energy surface projected along the
    SHAP CVs generated from the 3 site model
    grey lines characterize the combine density
    of the model and reference data and serve as a visual
    guide.}
    \label{fig:123}
\end{figure}

\begin{figure}
    \centering
    \centerline{\includegraphics[scale=0.6]{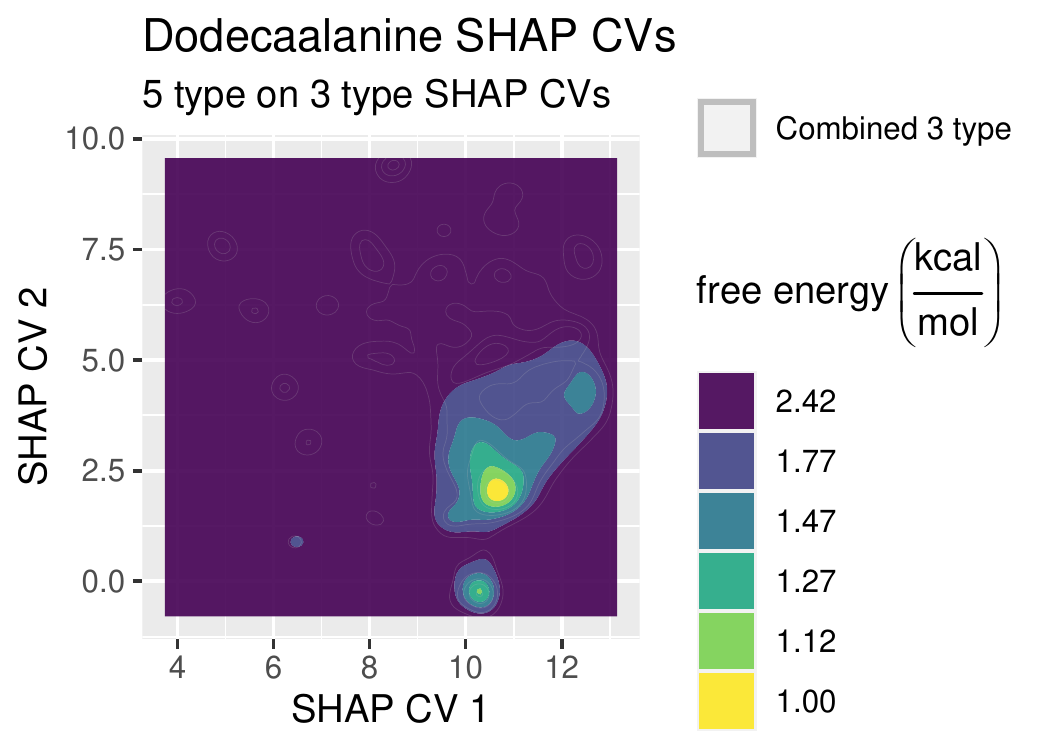}}
    \caption{5 type free energy surface projected along the
    SHAP CVs generated from the 3 type 
    model. Filled surfaces represent the 5 type model
    ensemble, while grey lines represent the combined 3 type
    and reference ensembles presented in Fig. 
    \ref{fig:123}.}
    \label{fig:123proj}
\end{figure}

In order to better describe the remaining error in the more
complex models, new SHAP CVs were generated based on the
comparison of the 3 type ensemble to the reference data
(Fig. \ref{fig:123}). 
The global geometry described by the resulting SHAP CVs
differs significantly from that present in Fig.
\ref{fig:nonb}: as expected, the bond errors dominating 
the 1 type model without nonbonded interactions
are no longer present. Inspection of the
individual SHAP values no longer isolates a particular
portion of the protein
as broadly responsible for $\deltau$ (see SI). The
large diffuse area present (as well as the protrusion at
$(10,0)$) in the model ensemble is due to erroneous
helix-like formation at
the C-terminus, a phenomena spread over 4 sites. The
subbasins near $(8,2)$, are due to differences in the inner
walls present in $\deltau$. 
Sharper features around $(4,6.5)$ are due to the remaining helix
formation (primarily present though 1-4 distances), while features 
around $(10,7.5)$
are due to a higher propensity for shorter bond
distances and angles throughout the middle of the protein.
These types of errors are degenerate along the protein
backbone, resulting in many minima in the surface. The
complex nature of the presented error makes it difficult to
isolate small adjustments to the force-field basis to
improve the model.
Attempting to improve the quality of the model by adding
additional site detail to each termini results in little
improvement, consistent with said error analysis. 
This is seen in both Fig. \ref{fig:ddaviolinerror} and
\ref{fig:123proj}, where Fig. \ref{fig:123proj} visualizes
the ensemble produced by the 5 type model projected
into the coordinates created from the 3 site model. 
This
lack of improvement contrasts strongly the improvement seen
in Fig. \ref{fig:nonbproj}, supporting the view that additional
parameters, if not targeted towards known areas of error,
do not necessarily improve model performance.

\subsection{Actin}\label{actinresults}
\begin{figure}
    \centering
    \centerline{\includegraphics[scale=0.6]{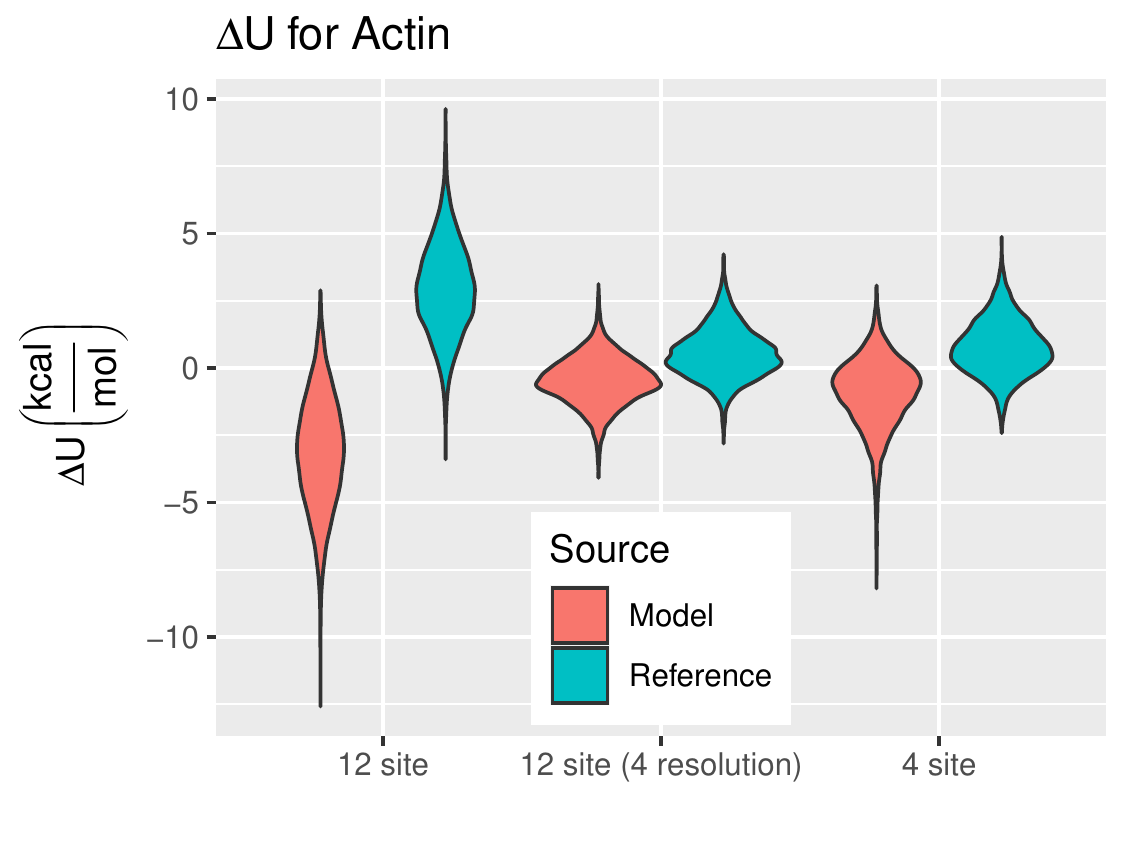}}
    \caption{Violin plots of 
    $\deltau$ for multiple models and resolutions of actin,
    divided along the reference 
    and model ensembles. The ``12 site (4
    resolution)'' and ``4 site'' models are compared to the
    reference at the 4 site resolution, while ``12
    site'' is compared at the 12 site resolution. Note that
    varying models/resolutions change the form of $\deltau$,
    which changes the shape of the projected reference
    distribution.}
    \label{fig:actinviolinerror}
\end{figure}

\begin{figure}
    \centering
    \centerline{\includegraphics[scale=0.6]{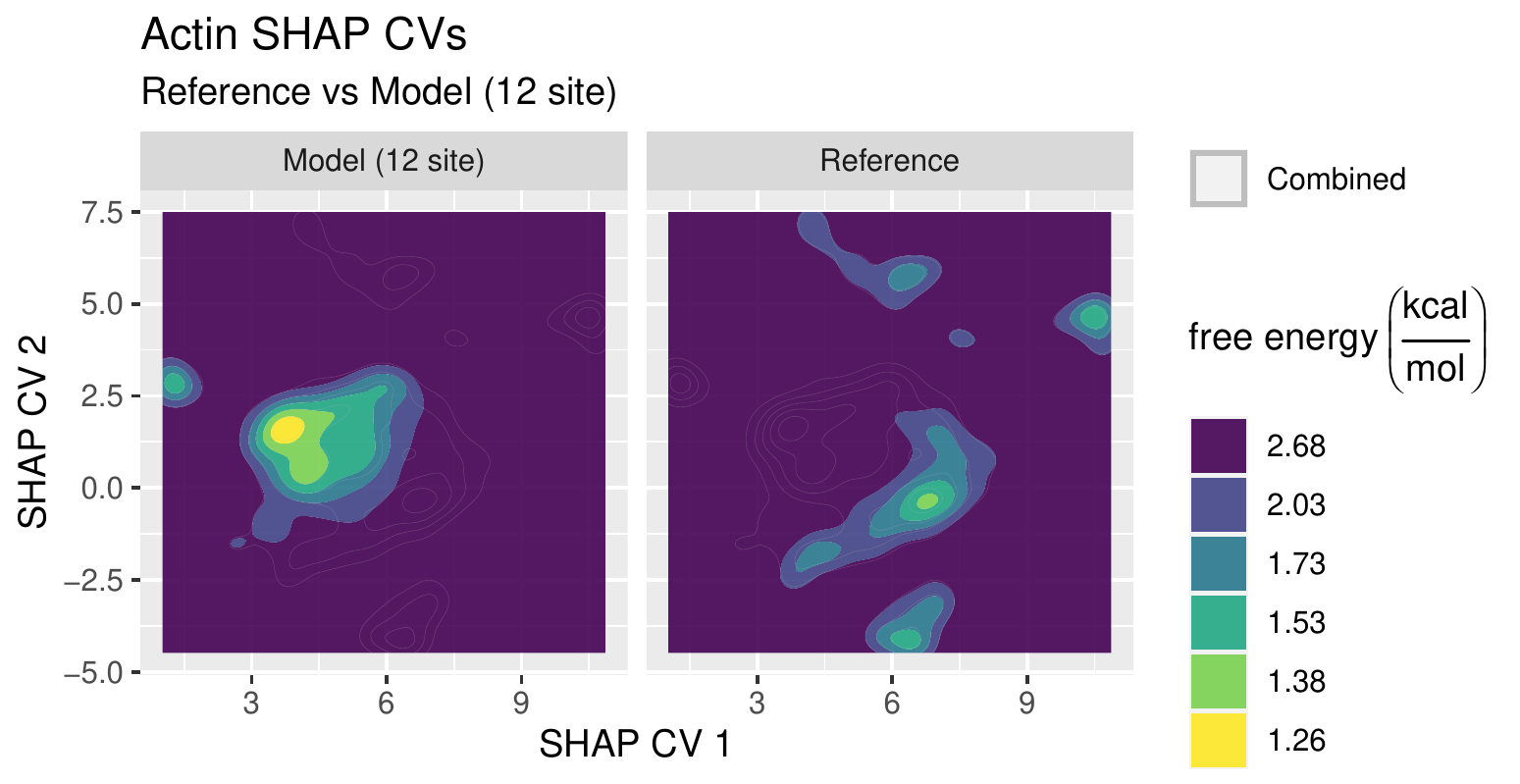}}
    \caption{The free energy plot given by the SHAP CVs
    calculated by comparing the 12 site elastic network to
    the reference ensemble at a 12 site resolution.}
    \label{fig:12actin}
\end{figure}

\begin{figure}
    \centering
    \centerline{\includegraphics[scale=0.6]{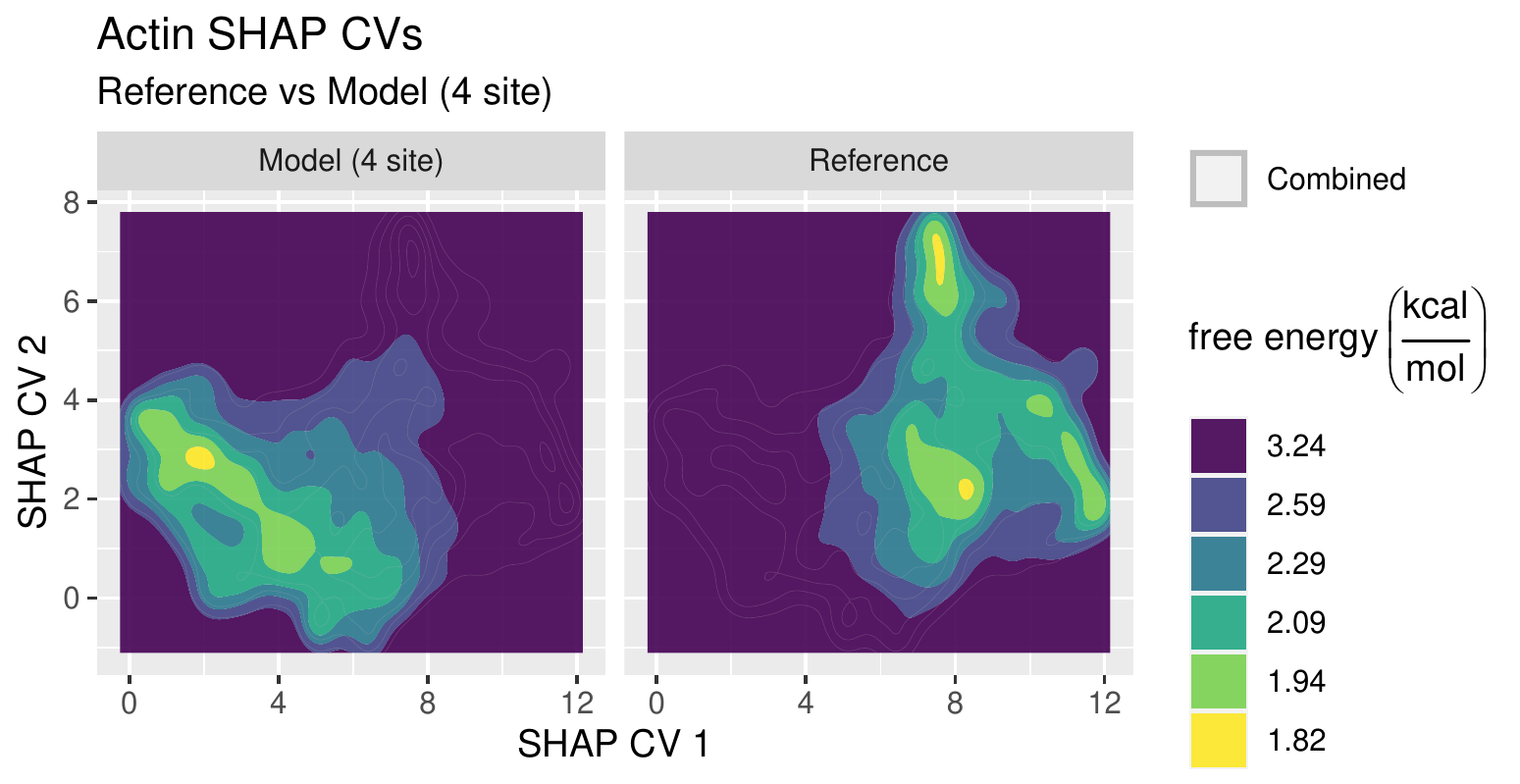}}
    \caption{Free energy surface produced along the SHAP
    variables generated by comparing a 4 site elastic
    network model to mapped reference data.}
    \label{fig:4actin}
\end{figure}

Actin was analyzed using two different elastic network
models created using the heteroENM 
methodology\cite{lyman2008systematic}: one model
was created at a 12 site resolution, while the other was
created using a 4 site subset of the 12 site model (sites
indexed 1 through 4). In order to
better understand model error, the samples produced by the
12 site model were additionally mapped to the 4 site
resolution for comparison to the 4 site model. The
distribution of $\deltau$ for the 12 site model at both
resolutions as well as the errors in the 4 site model are visualized in
Fig. \ref{fig:actinviolinerror}. At the 12 site resolution,
the 12 site model exhibits a large spread of $\deltau$. When
mapped to the 4 site resolution, however, it performs
marginally better
than the 4 site model when compared to the reference
ensemble.
SHAP variables were generated between the 12 site elastic
network and the reference ensemble (Fig. \ref{fig:12actin})
at the 12 site resolution; a
diffuse central basin is present, along with a variety of
smaller subbasins. The subbasins present at $(1.5,2.5)$ and
$(11,5)$ are due to the distance between sites 1 and 9. The
subbasin at $(6,5)$ is attributable to the distance between
sites 2 and 4. The subbasin at $(6,-3)$, along with the broadness of 
the central basin, correspond to relatively uncorrelated 
heterogeneity occurring between site 5 and the rest of the
protein, along with variation in the position of site 9. 
Collectively, the majority of
the error present is associated with sites $5$ and $9$,
which represent the d-loop (a transiently structured region
located on the edge of the protein)
and the nucleotide ADP (situated in the center of the
protein). This suggests that a model built at the
4 site resolution may result in less error, which is indeed shown
in Fig. \ref{fig:actinviolinerror}.
A harmonic network built at the 4 site resolution
was used to test this hypothesis; the resulting SHAP CV based 
comparison is shown
in Fig. \ref{fig:4actin}. 
Inspection of plotted SHAP values
indicates that the horizontal spread (along CV 1) is due to error
associated with the 2-4 and 3-4 distances, while the vertical
spread is attributable to error in the 1-4 distance. 

\begin{figure}
    \centering
    \centerline{\includegraphics[scale=0.6]{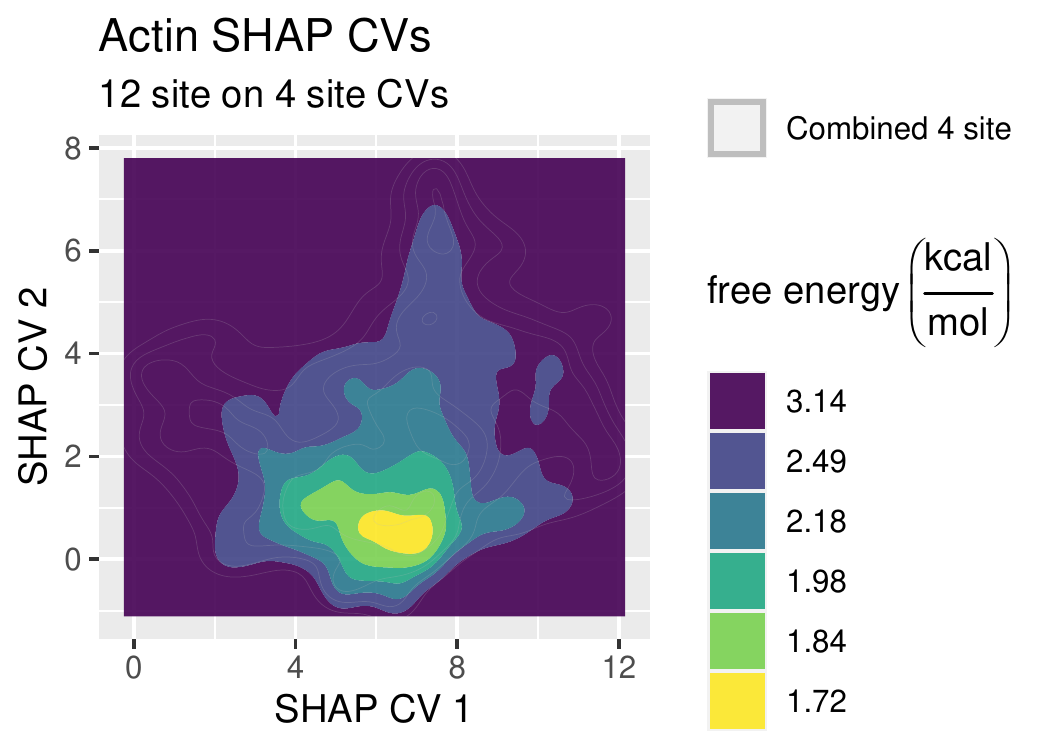}}
    \caption{12 site free energy surface produced along the SHAP
    variables generated by comparing a 4 site elastic
    network model to mapped reference data.
    Filled surfaces represent the 12 site model
    ensemble, while grey lines represent the combined 4
    site 
    model and reference ensembles presented in Fig. 
    \ref{fig:4actin}.
    }
    \label{fig:12actinat4proj}
\end{figure}

\begin{figure}
    \centering
    \centerline{\includegraphics[scale=0.6]{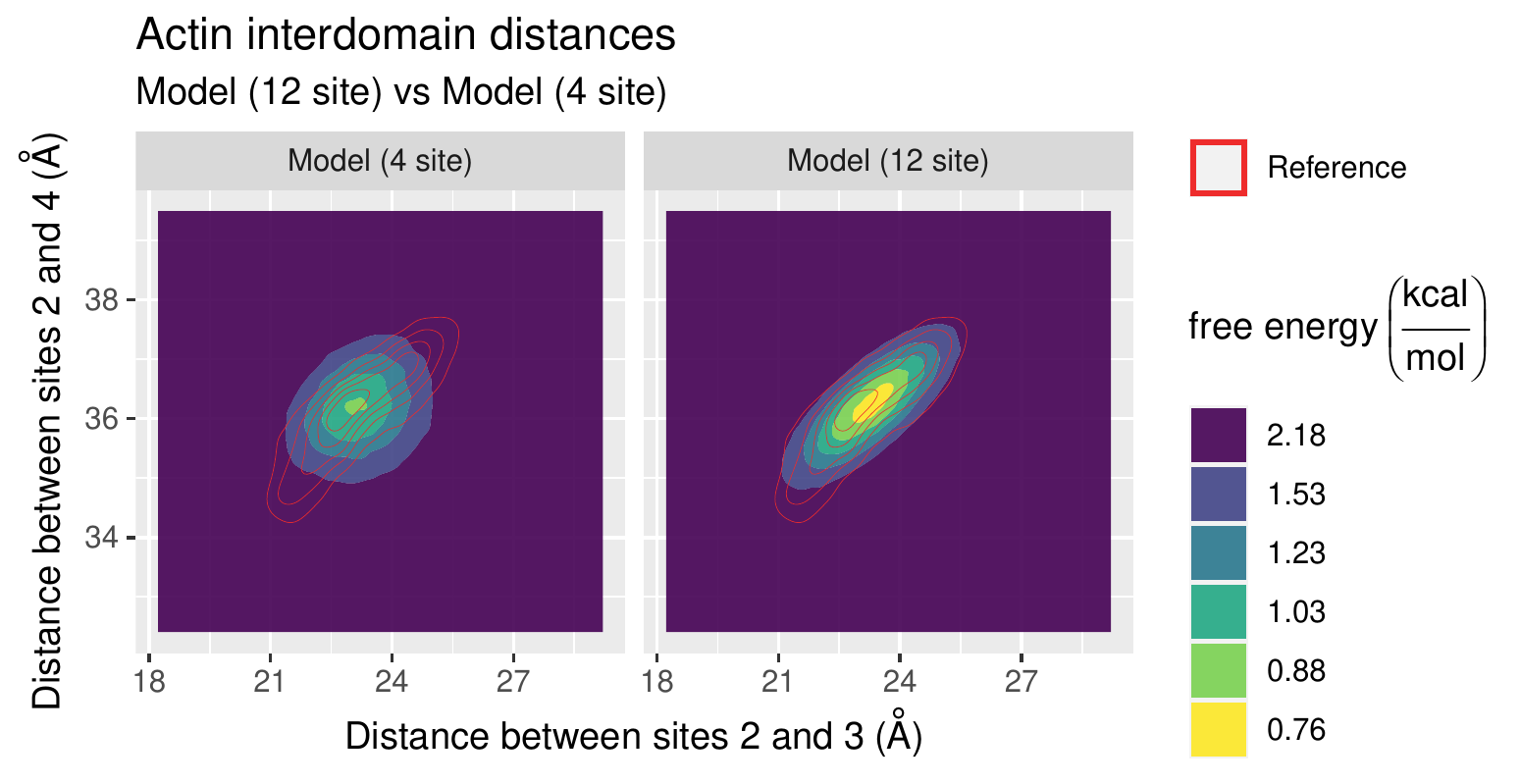}}
    \caption{Free energy surface produced the distance
    between sites 2 and 4 and the distance between sites 2
    and 3. Filled contour panels represent the two model
    ensembles, while the red line contour overlay correspond to
    statistics from the reference ensemble.}
    \label{fig:4vs12actinat4physical}
\end{figure}

However, $\deltau$ for the 4 site model is determined at a
different resolution than that of the 12 site model. As
where the 4 site model is only judged on the behavior of sites
1-4 (which represent the core of each of the 4 main
subdomains), the 12 site model is compared using all 12
sites--- and as a result must capture more complex behavior.
This likely does not align with error validation in
practice: if both a 4 site and 12 site model are considered,
it is likely that only sites 1-4 are critical (if not, the
4 site model would be an invalid choice regardless of its
accuracy). In other words, the pertinent behavior of the 12 site
model may very well be concentrated in sites 1-4. The
current analysis framework allows us to analyze the 12 site
model at the resolution of sites 1-4
by mapping the 12 site
model to the 4 site resolution. The resulting mapped 12 site
ensemble is then projected onto the variables generated by the
SHAP variable 4 site model (Fig. \ref{fig:12actinat4proj}).
The offset of the 12 site ensemble
in CV space (along with Fig. \ref{fig:actinviolinerror}) implies that 
errors related to the 2-4 and 2-3
are resolved in the 12 site model; this is reinforced by
considering the free energy surface given by said distances (Fig.
\ref{fig:4vs12actinat4physical}).

\section{Discussion and Conclusions}

The results presented demonstrate two uses of SHAP CVs.
First, the error of an existing model can be described and
used as a guide to adjust the force-field basis.  These CVs
can similarly be used to quantify the effect of a proposed
change to said basis.  Second, the effect of resolution on
the accuracy of a CG model can be quantified: in this case,
an increase in resolution improved the behavior of a CG
model quantified at a coarser resolution. While the
conclusions and model adjustments are not necessarily
counterintuitive, they were analyzed with minimal human
intervention.

In the case of DDA, the largest reduction in error was seen
when accounting for bond disagreements. This is supported by
the fact that the large free energies found in effective
bonded interactions can easily result in areas of phase
space that are effectively not traversed by either the model
or reference data, as a lack of overlap in any dimension
implies a lack of overlap at the full phase space
resolution. The conclusions drawn from the SHAP CVs contrast
strongly with conclusions drawn from pre-selected CVs: while
Fig. \ref{fig:rogqhel} suggests that the CG model
occupies a subset of the phase space favorable to the mapped
atomistic data, Fig. \ref{fig:ddaviolinerror}, along with
Figs \ref{fig:nonb} and \ref{fig:123}, illustrate that the
two ensembles are close to disjoint, while the latter two
figures still represent variation in Q-helicity (and to a
lesser extent, radius of gyration). This does not imply that
pre-selected CVs cannot correctly discover issues with
overlap--- $\deltau$ is a CV in itself and may be
selected/approximated using external knowledge.

In the case of actin, it is perhaps surprising that at a 12
site resolution the distribution of $\deltau$ is of a
similar magnitude as that found in DDA: actin has a
persistent tertiary structure while DDA is relatively
disordered. However, it is important to note that actin was
modeled using harmonic potentials as where DDA was modeled
using spline interactions. Based on SHAP based analysis, this error is
primarily due to difficulty modeling the d-loop and
nucleotide, a conclusion supported by the reduction of error
when considered at the 4 site resolution.  Notably, however,
the 12 site model performs better than one constructed
without these problematic sites due to improved reproduction
of cross correlations. This is attributable to the limited
basis of the forcefield, and suggests that approaches which
select CG mappings without considering the limited
force-field basis that is eventually applied will not
produce mapping operators that are necessarily useful in
practice. Similarly, if the proposed approach is combined
with a series of maps at varying resolutions, the effect of
model resolution can be validated in a novel and rigorous
way. The interplay between $\deltau$ and resolution
supports the idea that the correctness of an approximate CG model 
may be difficult to consider without an implied resolution
for analysis.

\subsection{Invariance and Adversarial Learning}
The relationship between $\cond$ and $\deltau$
(Eq. \eqref{eq:deltau}) provides important insight to
classification in the current context.  First, if the
force-fields generating both ensembles are known, the
information provided by classification can alternatively be gleaned by
calculating the overall difference in free energies using, for
example the Bennet Acceptance
Ratio\cite{bennett1976efficient}. Similarly, if
performing classification between a CG model and a mapped
reference ensemble, estimating the ideal classifier is
analogous to estimating the true PMF along with the
corresponding free energy difference.
Additionally, Eq. \eqref{eq:deltau} directly implies that any
symmetry or locality shared by $\ff$ and $\pmf$ is shared by
$\cond$. This has important consequences when considering
applying the proposed XAI approach to novel chemical
systems: the corresponding classification problem will
contain physical symmetries and locality, and approaches which
do not take this into account will likely provide poor
estimates of $\cond$. A straightforward example is a homogeneous liquid
where an asymmetric classifier is trained on the Cartesian coordinates:
sampling sufficient to converge various local
correlations (e.g., radial distribution functions) may be 
insufficient to parameterize such
a classifier. On the other hand, functions which are
formulated to obey permutational and rototranslational symmetries,
while widely investigated as techniques for atomistic and
CG force-field
development\cite{gkeka2020machine,goscinski2020role,
anderson2019cormorant}, will require
appropriate explanation methods.


Systematic coarse-graining methodologies typically define a
numerical measure of error and then return the force-field
which minimizes said error. Classification suggests similar
ideas of global error based on the accuracy achievable when
performing classification between the ensemble implied by
the force-field and the reference ensemble: a lower level of
mean accuracy implies better emulation of the reference
statistics (the accuracy is minimized if the reference and
model are indistinguishable; this results in a constant
$\cond$ of 0.5). A natural question is then to consider force-fields
which are optimized using this particular measure of quality. These
force-field optimization approaches lead to adversarial
learning, an approach firmly established by generative adversarial
networks\cite{goodfellow2014generative}, which when applied 
to CG force-field
development is termed Adversarial Residual Coarse-Graining
(ARCG)\cite{arcg}. 
The properties of $\cond$ described in the previous
paragraph additionally often apply to adversarial
learning\footnote{More precisely, they apply when the
variational process corresponds to classification. The
adversarial approach has since been expanded to encompass
additional divergences unrelated to classification.}. The
error estimation present in ARCG\cite{arcg} can resultingly be viewed
as simultaneously providing an estimate of $\pmf$ and 
the difference in configurational free energy. Conversely,
the variational error in ARCG can be calculated without
performing any classification: if a higher order force-field
is used to approximate $\pmf$ and supplemented with
a free energy difference method, the derivatives updating the
parameters are similarly calculable through
\eqref{eq:deltau}. Additionally, as $\cond$ is central to
adversarial residuals\cite{reid2011,arcg}, the explanations proposed in this article are
fundamentally related to global residuals such as relative
entropy and the Hellinger distance. The value of
these various divergences provide a quantification
of the overlap of $\deltau$ described Figs.
\ref{fig:ddaviolinerror} and \ref{fig:actinviolinerror};
however, the numerical values of these divergences are
difficult to interpret without context.

CG models are often created to study specific
phenomena, and it may not be necessary to perfectly produce
all the behavior of the mapped atomistic system. In this
cases the proposed methodology can be adapted by customizing
the resolution at which it is performed, such as in the
example of actin. However, more broadly, the concept of
bottom-up error analysis as presented here may not be
appropriate for these models. The modeler must decide
whether to view the model as a way to reproduce specific
phenomena or whether to view the model as a drop in
quantitative replacement for atomistic simulation. Certain
coarse-graining strategies, such as ARCG, can parameterize a
force-field to reproduce the manybody behavior of a subset
of the particles present in the CG system. However, doing so
incorporates additional human influence into the creation of
said CG model: as the resolution becomes coarser, the
approach begins to resemble top-down parameterization
strategies.  We note that machine learned atomistic
force-fields are often quantified including values similar to
$\deltau$; if CG models are to be eventually considered as
accurate as their fine-grained counterparts, utilizing
similar measures of quality is critical.

\subsection{XAI and future directions}
The analysis in this article focuses on using SHAP values to
describe the behavior produced by CG potentials.  The
approach trains a classifier to estimate $\deltau$, and then
uses techniques from XAI to explain said estimate.
Interpretable models and explanations intrinsically provide
a way to understand the high dimensional differences
characterizing the quality of a proposed CG
force-field, and we fully expect that other methods from the
rapidly developing field of XAI will find similar utility.
Furthermore, the study of explanations and interpetability
is fundamentally relevant to the creation of CG models: CG
force-fields are rarely created solely to reproduce the
manybody-PMF of the training ensemble. They are
instead often created to either extract knowledge from the
system under study or to investigate new physical settings,
tasks that intrinsically require human understanding of the
limitations and workings of the utilized CG model.  Any
technique for bottom-up CG model creation which uses
external human validation is a candidate for using
explainable techniques.  We hope that this work will serve
as an initial example for a new approach to CG model 
validation.

\subsection{References}
\begin{acknowledgement}
AEPD thanks Alexander Pak, Glen Hocky, and Sriramvignesh
Mani for insight, guidance, and simulation data. This
material is based upon the work supported by the
National Science Foundation (NSF, Grant No.
CHE-1465248). Simulations were performed using the
resources provided by the University of Chicago Research
Computing Center (RCC). 
\end{acknowledgement}
\bibliography{cg}
\end{document}